\documentclass[preprint,aps,12pt,preprintnumbers,eqsecnum,nofootinbib]{revtex4}
\usepackage{graphicx}
\usepackage{color}
\usepackage{amssymb,amsmath}

\begin{document}
%
%
\title{\vspace*{0.5in} A Froggatt-Nielsen Model for \\ Leptophilic Scalar Dark Matter Decay
\vskip 0.1in}
\author{Christopher D. Carone}\email[]{cdcaro@wm.edu}
\author{Reinard Primulando}\email[]{rprimulando@email.wm.edu}
\affiliation{High Energy Theory Group, Department of Physics,
College of William and Mary, Williamsburg, VA 23187-8795}
\date{May 2011}
\begin{abstract}
We construct a model of decaying, TeV-scale scalar dark matter motivated by data from the PAMELA and
Fermi-LAT experiments. By introducing an appropriate Abelian discrete symmetry and an intermediate
scale of vector-like states that are responsible for generating lepton Yukawa couplings, we show that
Planck-suppressed corrections may lead to decaying dark matter that is leptophilic and has the desired
lifetime.  The dark matter candidate decays primarily to lepton/anti-lepton pairs, and at a subleading
rate to final states with a lepton, anti-lepton and standard model Higgs boson.  We show that the model
can reproduce the observed positron flux and positron fraction while remaining consistent with the
bounds on the cosmic ray antiproton flux.
\end{abstract}
\pacs{}
\maketitle

\section{Introduction} \label{sec:intro}
A number of earth-, balloon-, and satellite-based experiments have observed anomalies in
the spectra of cosmic ray electrons and positrons. Fermi-LAT~\cite{Abdo:2009zk} and H.E.S.S.~\cite{Aharonian:2009ah} have measured an excess in the flux of electrons
and positrons up to, and beyond $1$~TeV, respectively.  PAMELA~\cite{Adriani:2008zr}, which
is sensitive to electrons and positrons up to a few hundred GeV in energy, detects an
upturn in the positron fraction beginning around 7 GeV, in disagreement
with the expected decline from secondary production mechanisms.  Recent measurements
at Fermi-LAT support this result~\cite{newfermi}. In contrast, current
experiments observe no excess in the proton or antiproton flux~\cite{Adriani:2008zq}.
Although astrophysical explanations are possible~\cite{astrox}, these observations can be
explained if the data includes a contribution from the decays of unstable dark matter
particles that populate the galactic halo~\cite{analyses}.  The dark matter candidate
must be TeV-scale in mass, have a lifetime of order $10^{26}$~seconds, and decay preferentially
to leptons. A number of scenarios have been proposed to explain the desired dark matter
lifetime and decay properties~\cite{CEP,anomaly,sgddm,models,nonab}.

To be more quantitative, consider a scalar dark matter candidate $\chi$ which (after the
breaking of all relevant gauge symmetries) has an effective coupling $g_{eff}$ to some standard
model fermion $f$ given by $g_{eff} \chi \bar f_L f_R + \mbox{h.c.}$  To obtain a lifetime of $10^{26}$~seconds, one finds $g_{eff} \sim 10^{-26}$ if $m_\chi \sim 3$~TeV.  From the perspective of naturalness, the origin of
such a small dimensionless number requires an explanation.  One possibility is that physics near the dark
matter mass scale is entirely responsible for the appearance of a small number, as is the case in models
where a global symmetry, that would otherwise stabilize the dark matter candidate, is broken by instanton
effects of a new non-Abelian gauge group $G_D$.  A leptophilic model of fermionic dark matter along these
lines was presented in Ref.~\cite{CEP}: the new gauge group is broken not far above the dark matter mass scale and the effective coupling is exponentially suppressed, $g_{eff} \propto \exp(-16\pi^2/g_D^2)$, where $g_D$ is the $G_D$ gauge coupling.  (An example of a supersymmetric model with anomaly-induced dark matter decays can be found in Ref.~\cite{anomaly}.) On the other hand, the appearance of a small effective coupling can arise if the breaking of the stabilizing symmetry is communicated to the dark matter via higher-dimension operators suppressed by some high scale $M$. Then it is possible that $g_{eff}$ is suppressed by $(m_\chi/M)^p$, for some power $p$; it is well known that for $m_\chi \sim {\cal O}(1)$~TeV and $p=2$, the correct lifetime can be obtained for $M \sim {\cal O}(10^{16})$~GeV, remarkably coincident with the grand unification (GUT) scale in
models with TeV-scale supersymmetry (SUSY)~\cite{sgddm}. If the LHC fails to find SUSY in the coming
years, however, then the association of $10^{16}$~GeV with a fundamental mass scale will no longer be strongly
preferred. Exploring other alternatives is well motivated from this perspective and, in any event, may provide valuable insight into the range of possible decaying dark matter scenarios.

The very naive estimate for $g_{eff}$ discussed above presumes that the result is determined by
a TeV-scale dark matter mass $m_\chi$, a single high scale $M$ and no small dimensionless factors.  Given
these assumption, the choice $M=M_*$, where $M_*=2 \times 10^{18}$~GeV is the reduced Planck mass, would
not be viable:  the dark matter decay rate is much too large for $p=1$ ({\em i.e.}, there would be no dark matter left at the present epoch) and is much too small for $p=2$ ({\em i.e.}, there would not be enough
events to explain the cosmic ray $e^\pm$ excess). However, Planck-suppressed effects arise so generically that
we should be careful not to discount them too quickly. What we show in the present paper is that
Planck-suppressed operators can lead to the desired dark matter lifetime if they correct new physics
at an intermediate scale.  In the model that we present, this is the scale at which Yukawa couplings
of the standard model charged leptons are generated via the integrating out of vector-like states.  This
sector will have the structure of a Froggatt-Nielsen model~\cite{fn}:  an Abelian discrete symmetry
will restrict the couplings of the standard model leptons and the vector-like states, but will be
spontaneously broken by the vacuum expectation values (vevs) of a set of scalar fields $\{\phi\}$.
Integrating out the heavy states will not only lead to the standard model charged lepton Yukawa couplings,
but also to dark matter couplings that are naturally leptophilic and lead to dark matter decay.   Aside
from setting the overall scale of the charged lepton masses, the symmetry structure of our
model will not restrict the detailed textures of the standard model Yukawa matrices.  This feature is
not automatic; symmetries introduced to guarantee dark matter leptophilia may also make it difficult
to obtain the correct lepton mass matrices, at least without additional theoretical
assumptions (for example, the addition of electroweak Higgs triplets, as in the model of Ref.~\cite{nonab}).
Our framework is free of such complications and is compatible, in principle, with many possible
extensions that might address the full flavor structure of the standard model.

Our paper is organized as follows.  In the next section, we present a model that illustrates
our proposal. In Section~3, we compute the predicted $e^\pm$ flux, $\Phi(e^\pm)$, and the positron
fraction $\Phi(e^+)/[\Phi(e^+)+\Phi(e^-)]$ for some points in the parameter space of our model and
compare our results to the relevant cosmic ray data. It is worth noting that this analysis has applicability to
any model that leads to similar dark matter decay operators.  In Section~4, we comment on the relic density and dark matter direct detection in our example model.  In Section~5, we summarize our conclusions.

\section{A Model}
We assume that the right-handed charged leptons of the standard model, $e_R$, and four sets of
heavy vector-like charged leptons are constrained by the discrete symmetry
\begin{equation}
G=\mathbb{Z}_p \times \mathbb{Z}_q \, ,
\end{equation}
with $p$ and $q$ to be determined shortly.  We assume that the vector-like leptons
have the same electroweak quantum numbers as $e_R$
\begin{equation}
E^{(i)}_R \sim E^{(i)}_L \sim e_R, \,\,\,\,\, (i=1 \ldots 4) \, .
\label{eq:eee}
\end{equation}
All the fields shown are assumed to be triplets in generation space, with their
generation indices suppressed.   Under the discrete symmetry, the fields in Eq.~(\ref{eq:eee})
are taken to transform as
\begin{equation}
e_R \rightarrow \omega^{-4} \, e_R \, ,
\label{eq:charges1}
\end{equation}
\begin{equation}
E_{L,R}^{(i)} \rightarrow \omega^{1-i} \, E_{L,R}^{(i)}, \,\,\,\,\, (i=1 \ldots 4)\, .
\label{eq:charges2}
\end{equation}
We will take $\omega$ and $\eta$ to be elements of $\mathbb{Z}_p$ and $\mathbb{Z}_q$, respectively, with $\omega^p=1$
and $\eta^q=1$.  In addition, we assume the presence of a heavy right-handed neutrino, $\nu_R$,
that is a singlet under $G$. We note that the fields that are charged under $G$ do not transform
under any of the non-Abelian standard model gauge group factors, so that $G$ satisfies the consistency
conditions of a discrete gauge symmetry in the low-energy theory~\cite{banksdine}; such discrete symmetries are
not violated by quantum gravitational effects\footnote{The consistency conditions require that anomalies involving
the non-Abelian gauge groups that are linear in a continuous group that embeds $G$ must vanish, as is automatic above.
Ref.~\cite{banksdine} indicates that no rigorous proof exists that the cancellation of the linear gravitational
anomalies is a necessary condition for the consistency of the low-energy theory.  Nonetheless, such a cancellation can
be achieved here by including a singlet, left-handed fermion, $N_L$, that transforms in the same way as $e_R$ under $G$.
For the choice $p=8$, adopted later in this section, $N_L$ can develop a Majorana mass somewhat below $M_*$ and
decay rapidly to lighter states via Planck-suppressed operators.  Including such a state does not
affect the phenomenology of the model otherwise.}.
The Yukawa couplings of the standard model charged leptons arise when the
symmetry $G$ is spontaneously broken and the vector-like leptons are integrated out of the
theory.  Symmetry breaking is accomplished via the vacuum expectation values of two scalar fields
$\phi_E$ and $\phi_D$, which transform as
\begin{eqnarray}
&& \phi_E \rightarrow \omega \, \phi_E \, , \nonumber \\
&& \phi_D \rightarrow \eta \, \phi_D \, .
\end{eqnarray}
The following renormalizable Lagrangian terms involving the charged lepton fields are allowed by
the discrete symmetry:
\begin{eqnarray}
{\cal L}_E &=& \overline{L}_L H E_R^{(1)} + \sum_{i=1}^3 \overline{E}^{(i)}_L \phi_E E^{(i+1)}_R
+ \overline{E}^{(4)}_L \phi_E \,e_R
\nonumber \\
&+& \sum_{i=1}^4 M^{(i)} \,\overline{E}^{(i)}_L E^{(i)}_R + \mbox{ h.c.}
\label{eq:esector}
\end{eqnarray}
While it is not our goal to produce a theory of flavor, we note that the terms in
Eq.~(\ref{eq:esector}) are of the type one expects in flavor models based on the
Froggatt-Nielsen mechanism.  Hence, integrating out the $E$ fields leads to a higher-dimension
operator
\begin{equation}
{\cal L} \supset \frac{1}{M^4} \overline{L}_L H \phi_E^4 e_R + \mbox{ h.c.} \, ,
\label{eq:chlep}
\end{equation}
which provides an origin for the charged lepton Yukawa couplings.  Choosing
$\langle \phi_E \rangle/M \sim 0.3$ gives the correct scale for the tau lepton
Yukawa coupling; the smaller, electron and muon Yukawa couplings may be accommodated
by suitable choices of the undetermined couplings in Eq.~(\ref{eq:esector}). One might
imagine that the remaining Yukawa hierarchies could be arranged by the imposition of
additional symmetries, though we will not explore that possibility here.

We now introduce  our dark matter candidate $\chi$, a complex scalar field that transforms
as
\begin{equation}
\chi \rightarrow \omega^4 \, \chi \,\,\,\,\, \mbox{ and } \,\,\,\,\,
\chi \rightarrow \eta^{-2} \chi \,
\label{eq:chicharges}
\end{equation}
under $\mathbb{Z}_p \times \mathbb{Z}_q$. We assume that all the nonvanishing powers of $\omega$ and $\eta$ shown in Eqs.~(\ref{eq:charges1}), (\ref{eq:charges2}) and (\ref{eq:chicharges}) are nontrivial, which requires
that $p>4$ and $q>2$.   Then, there are no renormalizable interactions involving a single $\chi$ field
(or its conjugate) and two fermionic fields that could lead to dark matter decay.  However, non-renormalizable,
Planck-suppressed operator provide the desired effect.  The lowest-order, Planck-suppressed correction to
Eqs.~(\ref{eq:esector}) that involves a single $\chi$ field is the unique dimension-six operator
\begin{equation}
\Delta {\cal L}_e = \frac{1}{M_*^2} \chi \, \overline{E}^{(1)}_L \phi_D^2 \, e_R + \mbox{ h.c.}
\label{eq:d6opE}
\end{equation}
Including Eq.~(\ref{eq:d6opE}) and again integrating out the heavy, vector-like states, one obtains
a new higher-dimension operator,
\begin{equation}
{\cal L}_{decay} = \frac{\phi_D^2}{M M_*^2}\, \chi \overline{L}_L H e_R + \mbox{ h.c.},
\label{eq:newop}
\end{equation}
which leads to dark matter decay.  For $m_\chi \sim 3$~TeV (compatible qualitatively with fits to the
PAMELA and Fermi-LAT data), a lifetime of $10^{26}$ seconds is obtained when
\begin{equation}
\frac{\langle \phi_D \rangle^2}{M_*^2}\frac{\langle H \rangle}{M} \sim 1 \times 10^{-26} \,.
\end{equation}
For our operator expansion to be sensible, we require $\langle \phi_D \rangle < M$; however, we also do not
want a proliferation of wildly dissimilar physical scales, if this can be avoided. Interestingly, if we choose $M$ to
be the geometric mean of $\langle H \rangle$ and $M_*$, one finds
\begin{equation}
M = 2 \times 10^{10}\mbox{ GeV}, \,\,\,\,\,
\langle \phi_E \rangle = 0.3\, M, \,\,\,\,\,
\langle \phi_D \rangle =0.1\, M\,,
\end{equation}
which meets our aesthetic requirements. Standard model quark and neutral lepton masses are unaffected by the discrete symmetry of our model, by construction. Light neutrino masses arise via a conventional see-saw
mechanism, and it is possible to obtain a right-handed neutrino mass scale $M_R \approx M$, so that all the heavy leptons appear at a comparable scale.  Assuming that the largest neutrino squared mass is comparable to $\Delta m^2_{32}=2.43 \times 10^{-3}$~eV$^2$, as suggested by atmospheric neutrino oscillations~\cite{pdg}, then this possibility is obtained if the overall scale of the Yukawa coupling matrix that appears in the neutrino Dirac mass term is of the same order as the charm quark Yukawa coupling.
\begin{figure}
    \centering
        \includegraphics[width=8cm,angle=0]{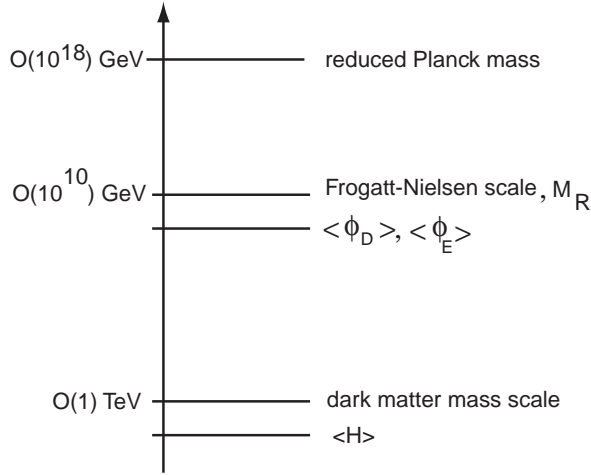}
    \caption{A possible choice for the mass scales in the theory. Symmetry breaking
    vevs appear within approximately an order of magnitude of the lower two scales.}
    \label{fig:scales}
\end{figure}
This scenario is depicted in Fig.~\ref{fig:scales}. In this case, the theory is characterized
by three fundamental scales: the Planck scale, an intermediate scale (associated with charged lepton flavor
and right-handed neutrino masses), and the TeV-scale. Symmetry-breaking vevs appear within a factor of $\alt 10$ below the latter two. Of course, the right-handed neutrino scale need not be linked with the scale at which
the charged lepton Yukawa couplings are generated; this is simply one of many viable possibilities that depend
on choices of the free parameters of the model.

Finally, we return to the discrete symmetry group $G=\mathbb{Z}_p \times \mathbb{Z}_q$.  We have noted that the structure of the theory that we have described is obtained for $p>4$ and $q>2$, but this does not take into account an important additional constraint:  there must be no Planck-suppressed operators involving couplings between the various scalar fields in the theory that can lead to other dark matter decay channels that are
either (i) too fast or (ii) too hadronic.  For example, the choice $p=5$ and $q=3$, allows the
renormalizable $G$-invariant operator  $\chi \phi_E \phi_D^\dagger$, which leads to mixing, for example,
between the $\chi$ and $\phi_E$ fields; the latter couples to two standard model leptons via the operator in
Eq.~(\ref{eq:chlep}), leading to a disastrously large decay rate.  We find that all unwanted operators are
sufficiently suppressed if we take $p=8$ and $q=4$, that is
\begin{equation}
G_I = \mathbb{Z}_8 \times \mathbb{Z}_4 \, .
\end{equation}
The lowest-order combination of scalar fields that is invariant under $G_I$, as well as the standard
model gauge group, is
\begin{equation}
\frac{1}{M_*^3} \chi \, \phi_D^2 \, \phi_E^4 \, ,
\label{eq:thecombo}
\end{equation}
Suppression by {\em three} factors of the Planck scale is more than sufficient to suppress any operators that
are generated when the $\phi_E$ and $\phi_D$ fields are integrated out of the theory, or that may be constructed from products of Eq.~(\ref{eq:thecombo}) with any $G_I$-singlet, gauge-invariant combination of standard model
fields. It is straightforward to confirm that the alternative choice
\begin{equation}
G_{II}=\mathbb{Z}_8 \times \mathbb{Z}_5  \, ,
\end{equation}
is also viable, by similar arguments.  The difference between the symmetry groups $G_I$ and $G_{II}$ is that
the former allows two types of dark matter mass terms: $\chi^2 + \mbox{h.c.}$ and $\chi^\dagger \chi$.  This
leads to a mass splitting between the two real scalar components of $\chi$, so that the lighter is the dark
matter candidate.  The choice $G_{II}$ forbids the $\chi^2$ mass terms, so that the dark matter consists of
particles and anti-particles associated with the original complex scalar field.  We note that in this theory,
the renormalizable interactions involving $\chi$ have an accidental U(1)$_\chi$ global symmetry which would lead to dark matter stability in the absence of the Planck-suppressed effects. The analysis that we present
in the following sections is somewhat simplified by the choice of $G_{II}$, which we adopt henceforth.

\section{Cosmic Ray Spectra}
\label{sec:positron}
In this section, we investigate the cosmic ray $e^\pm$ and proton/antiproton spectra of our model.
Our treatment of cosmic ray propagation follows that of Ref.~\cite{Ibarra:2009dr}.
We show that model parameters may be chosen to accommodate the positron excess and the
rising electron-positron flux observed by the PAMELA and Fermi-LAT experiments, respectively.

In Eq.~(\ref{eq:newop}), we identified the operator responsible for dark matter decays.  More
explicitly, this operator may be written
\begin{equation}
{\cal L}_{decay} = c_{ij} \frac{\langle \phi_D \rangle^2}{M M_*^2}\, \chi \overline{L}^i_L H e^j_R + \mbox{ h.c.},
\label{eq:opagain}
\end{equation}
where $i$ and $j$ are generation indices, and $c_{ij}$ represents unknown order-one coefficients.
Different choices for the couplings $c_{ij}$ will lead, in principle, to different cosmic ray
spectra.  To simplify the analysis, we focus on two possibilities: In the lepton mass eigenstate
basis, the fermions appearing in the decay operators are either ({\em i}) muons exclusively, or
({\em ii}) taus exclusively.  We will find that either of these choices is consistent with the data, even
though we have not fully exploited the parametric freedom available in the $c_{ij}$.  This is sufficient
to demonstrate the viability of our model. The remaining factors in the operator coefficient are chosen to
obtain the desired dark matter lifetime, as we discussed in the previous section.

In unitary gauge, the operator (\ref{eq:opagain}) can be be expanded
\begin{equation}
{\cal L}_{decay} = \frac{1}{\sqrt{2}} g_{ij} (v_{ew}+h) \, \chi \overline{e}_L^i \, e_R^j + \mbox{ h.c.},
\end{equation}
where $h$ is the standard model Higgs field, which we will assume has a mass of $117$~GeV, $v_{ew} = 246$~GeV,
and $g_{ij} \equiv c_{ij} \langle \phi_D \rangle^2 / (M M_*^2)$. The term proportional to the Higgs vev leads to the two-body decay $\chi \rightarrow \ell^+ \ell^-$, for $\ell=\mu$ or $\tau$, while the remaining term contributes to $\chi \rightarrow \ell^+ \ell^- h$.  We take both of these decay channels into account in our numerical analysis.  The final state particles in these primary decays will subsequently decay.  The electrons, positrons, protons and antiprotons that are produced must be added to expected astrophysical backgrounds to predict the spectra at experiments like PAMELA and Fermi-LAT.

Electrons and positrons that are produced in dark matter decays must propagate through the Milky Way before
reaching the Earth. In order to determine the observed fluxes, one must model this propagation.  The transport
equation for electron and positrons is given by
\begin{equation} \label{eq:transport}
0 = \nabla \cdot \left[ K(E,\vec r) \nabla f_{e^\pm}\right] +
\frac{\partial}{\partial E}\left [b(E,\vec r)f_{e^\pm}\right ] + Q_{e^\pm} (E,\vec r),
\end{equation}
where $f_{e^\pm} (E, \vec r, t)$ is the number density of electron or positrons per unit energy,
$K(E,\vec r)$ is the diffusion coefficient and $b(E,\vec r)$ is the energy loss rate.  We assume
the MED propagation model described in Ref.~\cite{Delahaye:2007fr}.  The diffusion coefficient and
the energy loss rate are assumed to be spatially constant throughout the diffusion zone and are given by
\begin{equation}
K(E, \vec r) = 0.0112 \epsilon^{0.70} \textrm{ kpc}^2/\textrm{Myr}
\end{equation}
and
\begin{equation}
b(E, \vec r) = 10^{-26} \epsilon^2 \textrm{ GeV/s} \, ,
\end{equation}
where $\epsilon = E/1$~GeV. The last term in Eq.~(\ref{eq:transport}) is the source term given by
\begin{equation} \label{eq:source}
Q(E,\vec r) = \frac{\rho(\vec r)}{M_\chi \tau_\chi} \frac{dN}{dE},
\end{equation}
where $M_\chi$ is the dark matter mass and $\tau_\chi$ is the dark matter lifetime. In models like ours,
where the dark matter can decay via more than one channel, the energy spectrum $dN/dE$ is given by
\begin{equation}
\frac{dN}{dE} = \sum_i \frac{\Gamma_i}{\Gamma} \left( \frac{dN}{dE}\right)_i,
\end{equation}
where $\Gamma_i / \Gamma$ is the branching fraction and $(dN/dE)_i$ is the electron-positron energy
spectrum of the $i^{\rm th}$ decay channel.  We use PYTHIA~\cite{Sjostrand:2007gs} to determine the
$(dN/dE)_i$. For the dark matter density, $\rho(\vec r)$, we adopt the spherically symmetric
Navarro-Frenk-White halo density profile~\cite{Navarro:1995iw}
\begin{equation}
\rho(r) = \frac{\rho_0}{(r/r_c)[1+(r/r_c)]^2} \, ,
\end{equation}
with $\rho_0 \simeq 0.26 \textrm{ GeV/cm}^3$ and $r_c \simeq 20 \textrm{ kpc}$.  The solutions to the
transport equation are subject to the boundary condition $f_{e^\pm}=0$ at the edge of the diffusion zone,
a cylinder of half-height $L = 4$ kpc and radius $R = 20$ kpc measured from the galactic center.

The solution of the transport equation can be written
\begin{equation}
f_{e^\pm}(E) = \frac{1}{M_\chi \tau_\chi} \int_{0}^{M_\chi} dE' G_{e^\pm} (E,E') \frac{dN_{e^\pm} (E')}{dE'},
\end{equation}
where $G_{e^\pm} (E,E')$ is a Green's function, whose explicit form can be found in Ref.~\cite{Ibarra:2008qg}.
The interstellar flux then follows immediately from
\begin{equation}
\Phi^{DM}_{e^\pm} = \frac{c}{4\pi} f_{e^\pm}(E).
\end{equation}
We adopt a parameterization of the interstellar background fluxes given in Ref.~\cite{Ibarra:2009dr}:
\begin{equation}
\Phi_{e^-}^{bkg}(E) = \left( \frac{82.0\epsilon^{-0.28}}{1+0.224\epsilon^{2.93}} \right)
\textrm{ GeV}^{-1}\textrm{m}^{-2}\textrm{s}^{-1}\textrm{sr}^{-1},
\end{equation}
\begin{equation}
\Phi_{e^+}^{bkg}(E) = \left( \frac{38.4\epsilon^{-4.78}}{1+0.0002\epsilon^{5.63}}
+24.0\epsilon^{-3.41} \right) \textrm{ GeV}^{-1}\textrm{m}^{-2}\textrm{s}^{-1}\textrm{sr}^{-1}.
\end{equation}
Finally, the flux at the top of the earth's atmosphere, $\Phi_{e^\pm}^{TOA}$, is corrected by solar modulation effects~\cite{Ibarra:2009dr},
\begin{equation}
\Phi_{e^\pm}^{TOA} (E_{TOA}) = \frac{E_{TOA}^2}{E^2_{IS}} \Phi_{e^\pm}^{IS} (E_{IS}) \, ,
\end{equation}
where $E_{IS} = E_{TOA} + |e| \phi$, and $|e| \phi = 550$~MeV. $E_{IS}$ and $E_{TOA}$ are the energy of positron/electron at the heliospheric boundary and at the top of atmosphere, respectively.

The total electron and positron flux is determined by
\begin{equation}
\Phi^{tot} (E) = \Phi^{DM}_{e^-}(E) + \Phi^{DM}_{e^+}(E) + k \Phi^{bkg}_{e^-}(E) + \Phi^{bkg}_{e^+}(E) ,
\end{equation}
where $k$ is a free parameter that determines the normalization of the primary electron flux background. The positron excess is given by
\begin{equation}
PF(E) = \frac{ \Phi^{DM}_{e^+}(E) + \Phi^{bkg}_{e^+}(E) }{\Phi^{tot} (E)}.
\end{equation}

\begin{figure}
    \centering
        \includegraphics[width=8cm]{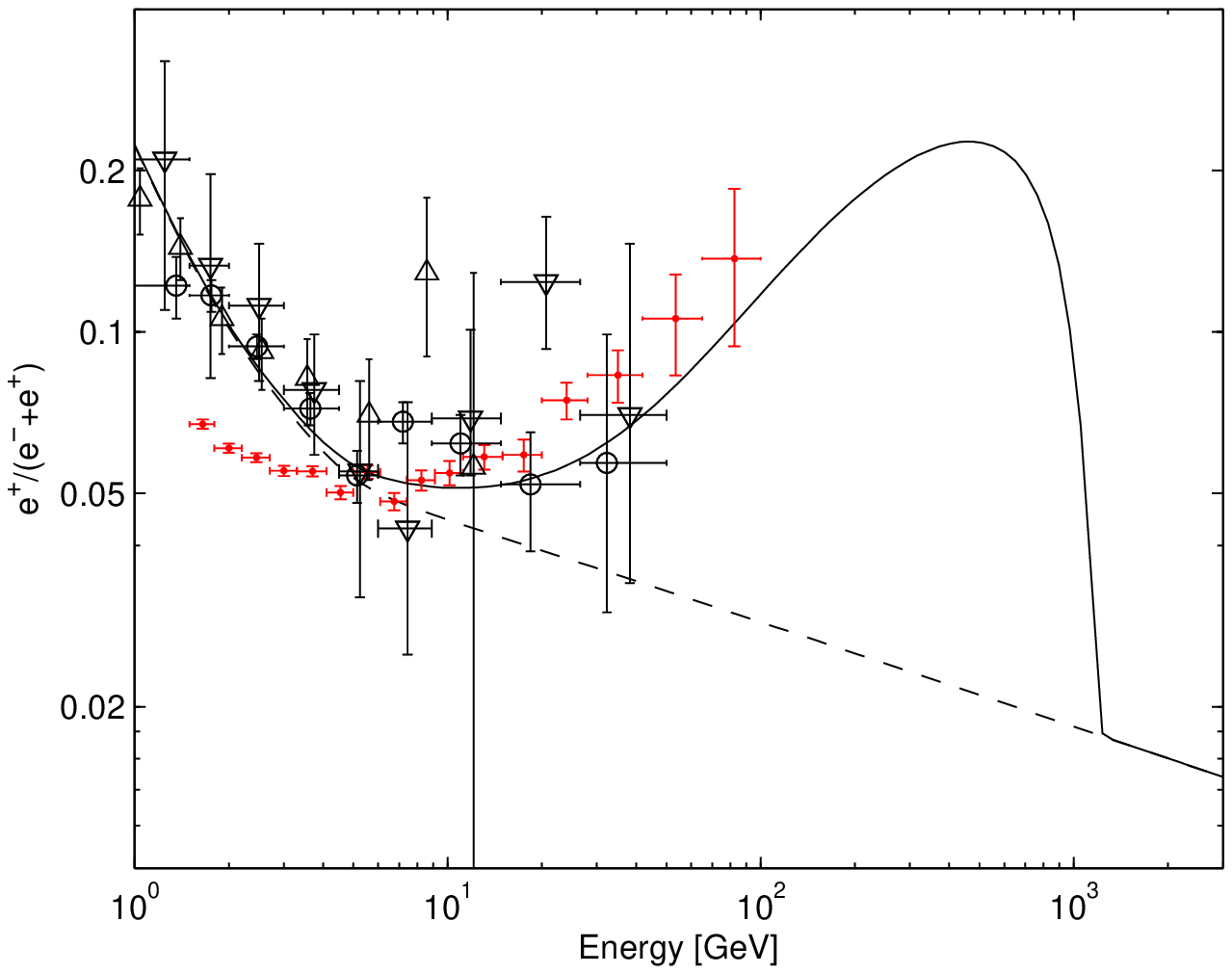}
        \includegraphics[width=8cm]{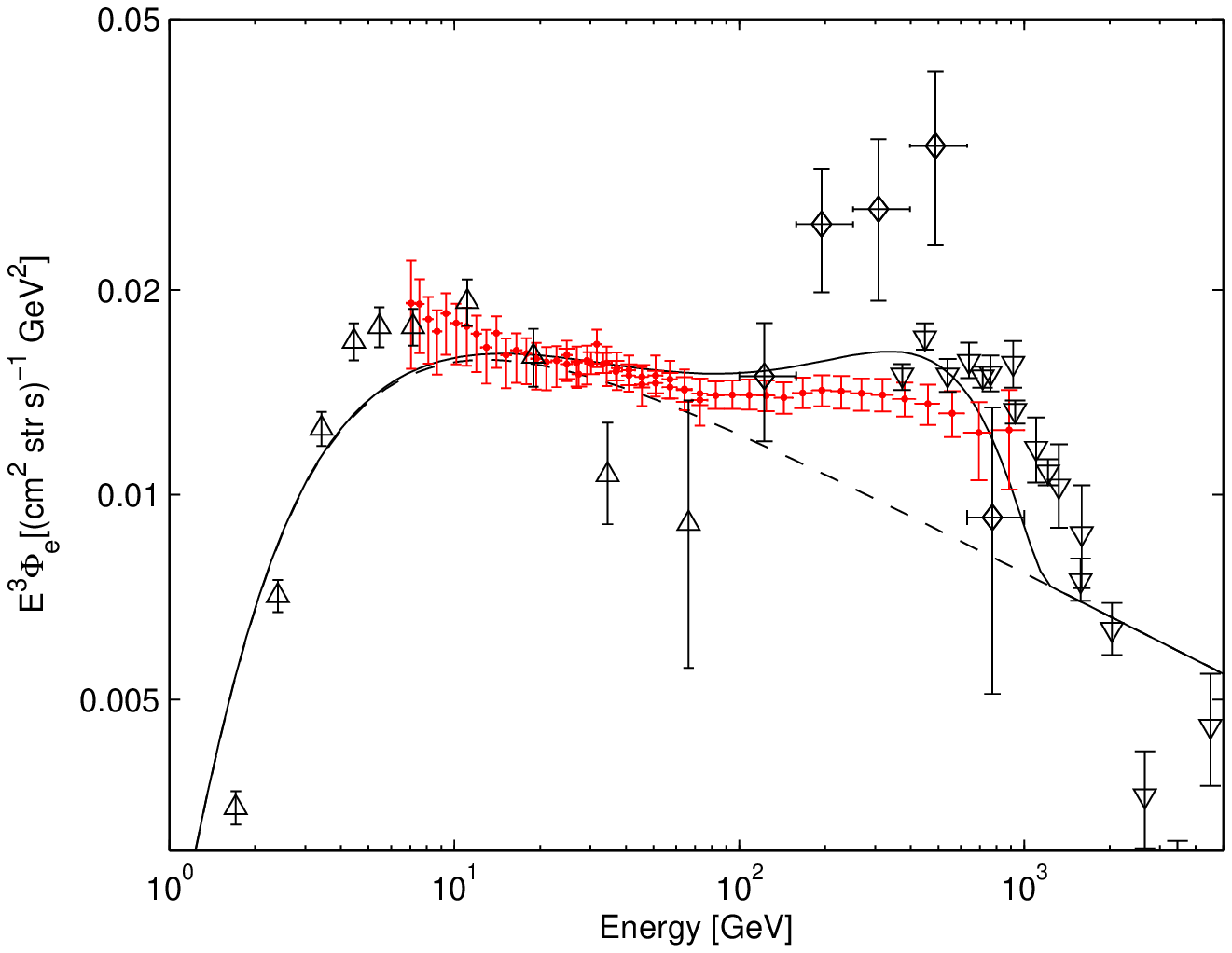}
    \caption{\textit{Left panel}: The positron excess for dark matter decaying into $\mu^+\mu^-$
    and $\mu^+\mu^-h$.  The dark matter mass is $2.5$~TeV and lifetime $1.8 \times 10^{26}$~s; the
    branching fraction to the two-body decay mode is $90.2\%$. The dashed line represents the background and the
    solid line represents the background plus dark matter signal.  Data from the following experiments are shown:
    PAMELA~\cite{Adriani:2008zr} (solid dots), HEAT~\cite{Barwick:1997ig} ($\circ$),
    AMS-01~\cite{Aguilar:2007yf} ($\bigtriangledown$), and CAPRICE~\cite{Boezio} ($\bigtriangleup$).
    \textit{Right panel}: The corresponding graph for the total electron and positron flux. Data from the following
    experiments are shown: Fermi-LAT~\cite{Ackermann:2010ij} (solid dots), HESS~\cite{Aharonian:2008aa} ($\bigtriangledown$),
    PPB-BETS~\cite{Torii:2008xu} ($\diamond$), HEAT~\cite{DuVernois:2001bb} ($\bigtriangleup$).}
    \label{fig:positronmu}
\end{figure}

\begin{figure}
    \centering
        \includegraphics[width=8cm]{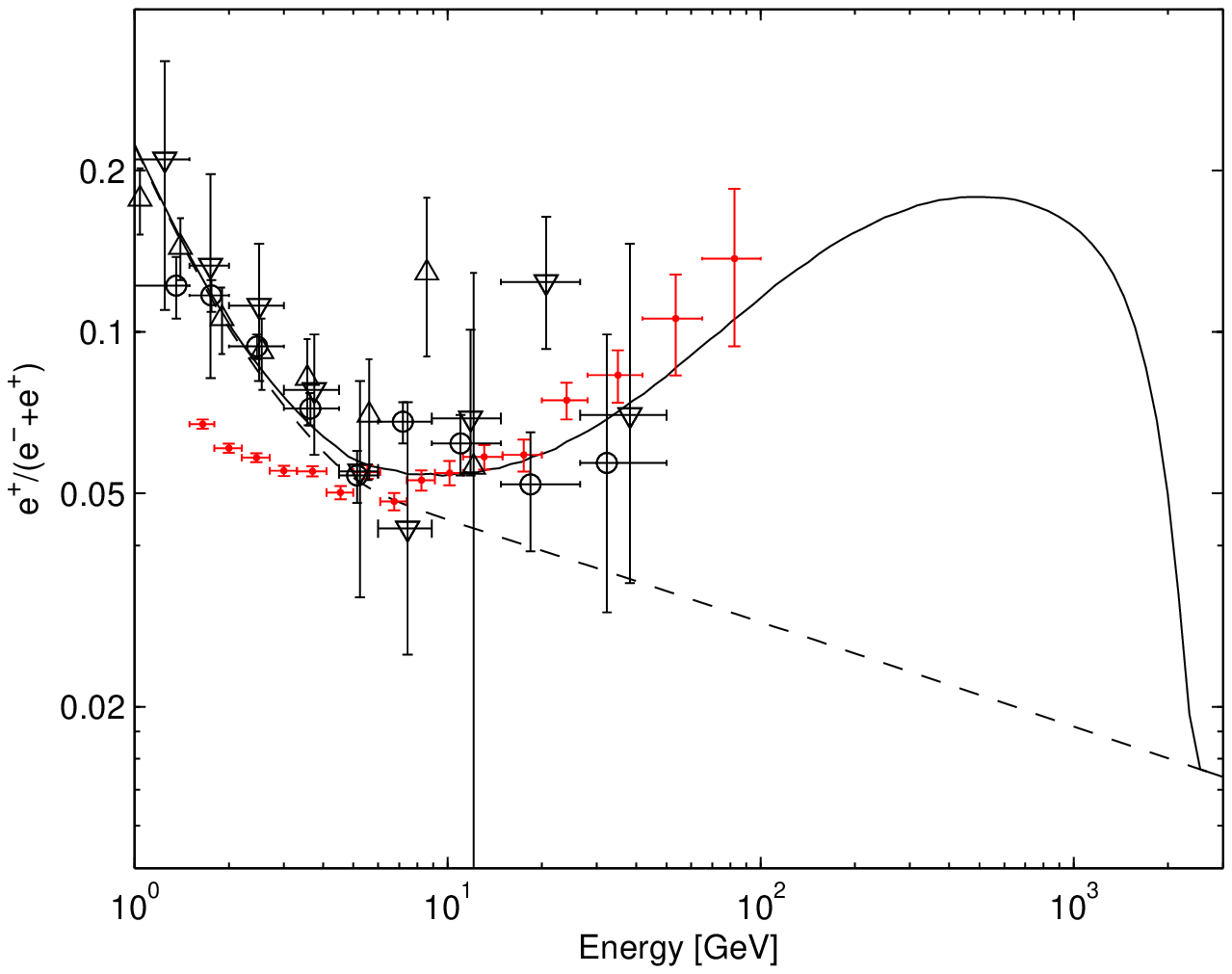}
        \includegraphics[width=8cm]{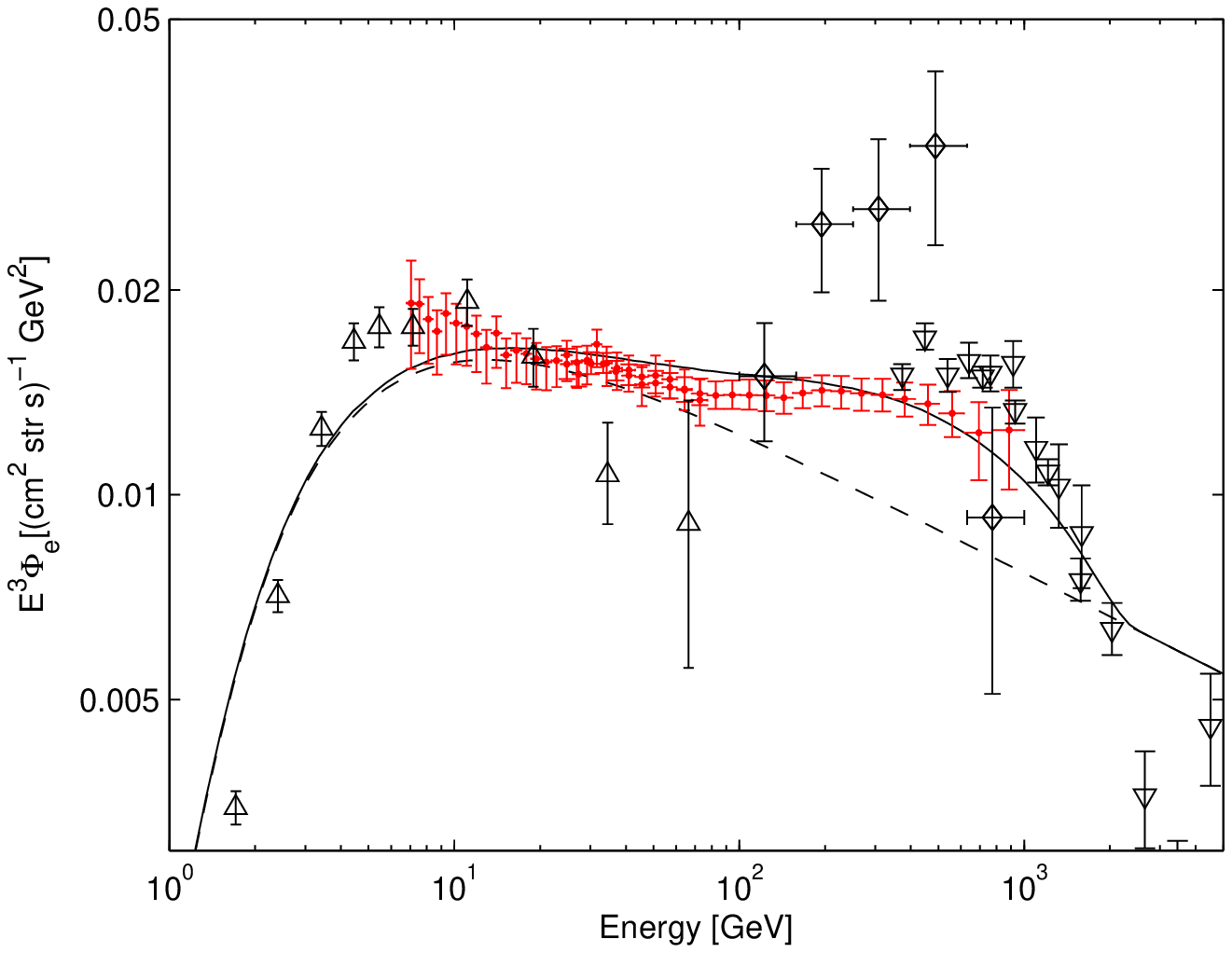}
    \caption{\textit{Left panel}: The positron excess for dark matter decaying into $\tau^-\tau^+$ and
    $\tau^-\tau^-h$. The dark matter mass is $5.0$~TeV and lifetime $9.0 \times 10^{25}$~s; the branching fraction
    to the two-body decay mode is 69.6\% . \textit{Right panel}: The corresponding graph for the total electron and
    positron flux.}
    \label{fig:positrontau}
\end{figure}

The results of our analysis are presented in Figs.~\ref{fig:positronmu} and \ref{fig:positrontau}.  In
the case where the dark matter decays only to $\mu^+\mu^-$ and $\mu^+\mu^-h$, we find good agreement with the
data for $\tau_\chi=1.8\times 10^{26}$~s and $M_\chi=2.5$~TeV.  In this case, the branching fraction
to the two-body decay mode is $90.2\%$. In the case where the decay is to $\tau^+\tau^-$ and $\tau^+\tau^-h$ only, our best results are obtained for $\tau_\chi=9.0 \times 10^{25}$~s and $M_\chi=5$~TeV, corresponding to
a two-body branching fraction of 69.6\%.  In all these results, the background electron flux parameter $k$ is set to $0.88$, following Ref.~\cite{Ibarra:2008qg}.

Since the dark matter decays in our model include the production of standard model Higgs bosons in the final state,
it is worthwhile to check that subsequent Higgs decays do not lead to an excess of cosmic ray antiprotons, in conflict
with the experimental data.  This will not be the case at our two benchmark parameter choices since the branching
fraction to the three-body decay mode is suppressed compared to the two-body mode.
The procedure for computing the cosmic ray antiproton flux is similar to that of the cosmic ray electrons and
positrons. The transport equation for antiproton propagation within the Milky Way is given by
\begin{equation}
0 = \nabla \cdot \left[ K(T,\vec r)\nabla f_{\bar p} - \vec V_c(\vec r) f_{\bar p} \right] + Q_{\bar p} (T,\vec r)
\end{equation}
where $T$ is the antiproton kinetic energy, $\vec V_c(\vec r)$ is the convection velocity, and the source term $Q_{\bar p}$ has the same form as Eq.~(\ref{eq:source}). As in the case of $e^\pm$ propagation, the antiproton number density can be expressed in terms of a Green's function
\begin{equation}
f_{\bar p} (T) = \frac{1}{M_\chi \tau_\chi} \int_0^{T_{max}} dT' G_{\bar p} (T, T') \frac{dN_{\bar p} (T')}{dT'},
\end{equation}
where $G_{\bar p}(T,T')$ can be found in Ref.~\cite{Ibarra:2008qg}. The relation between the antiproton number
density and the interstellar flux of antiproton is given by
\begin{equation}
\Phi_{\bar p}^{DM} (T) = \frac{v}{4\pi} f_{\bar p} (T) \, ,
\end{equation}
where $v$ is the antiproton velocity. We also take account the solar modulation effect on the antiproton flux at the top of atmosphere, $\Phi_{\bar p}^{TOA}$, which is given by
\begin{equation}
\Phi_{\bar p}^{TOA} (T_{TOA}) = \left( \frac{2m_pT_{TOA} + T_{TOA}^2}{2m_pT_{IS} + T_{IS}^2} \right)
\Phi_{\bar p}^{IS} (T_{IS}),
\end{equation}
where $T_{IS}$ and $T_{TOA}$ are the antiproton kinetic energies at the heliospheric boundary and at the top of atmosphere, respectively, with $T_{IS}= T_{TOA} + |e| \phi$. For the proton and antiproton flux, we adopt the background given in Ref.~\cite{Ptuskin:2005ax}.

\begin{figure}
    \centering
        \includegraphics[width=8cm]{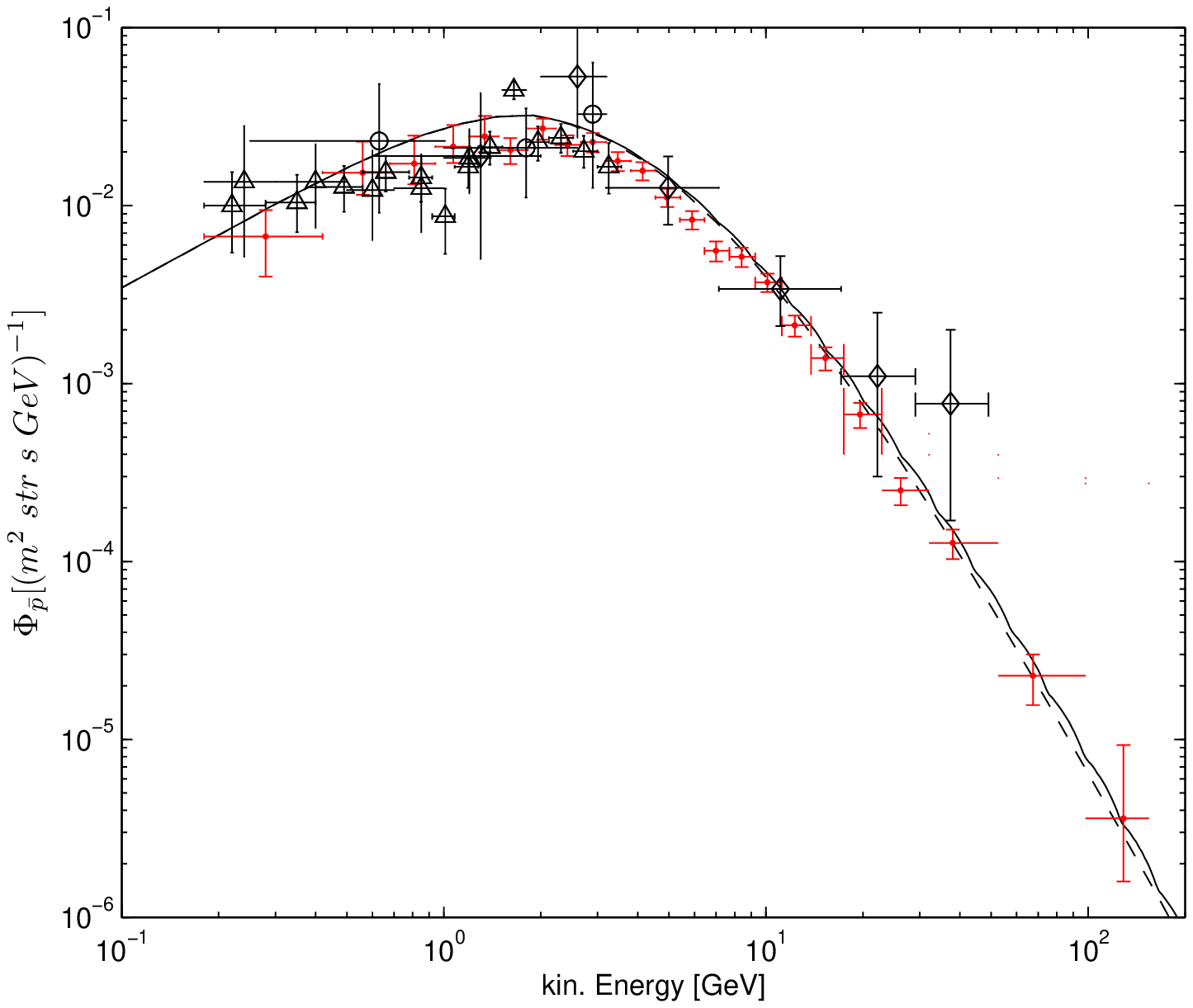}
        \includegraphics[width=8cm]{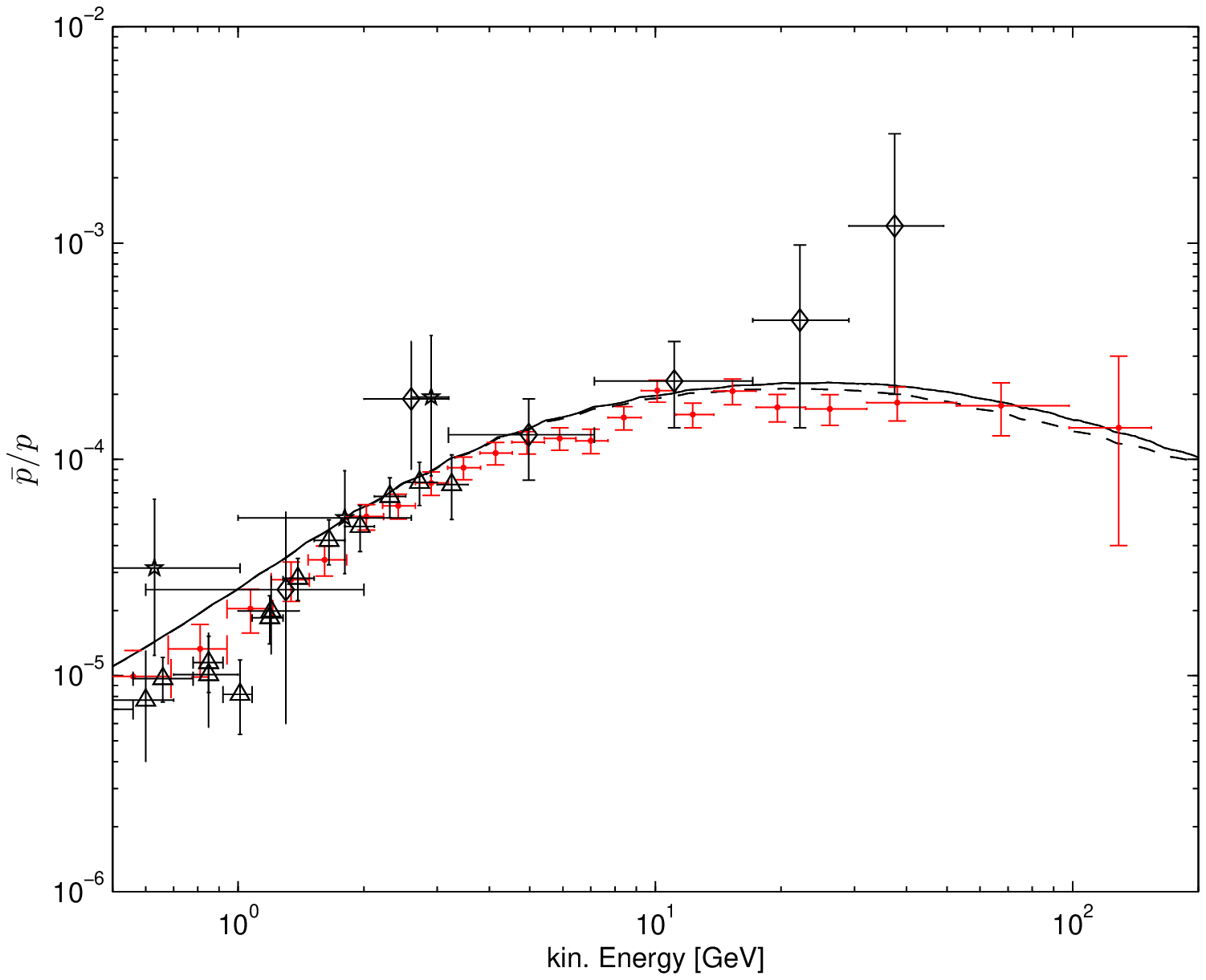}
    \caption{\textit{Left panel}:
The antiproton flux for dark matter decaying into $\mu^+\mu^-$
    and $\mu^+\mu^-h$.  The dark matter mass is $2.5$~TeV and lifetime $1.8 \times 10^{26}$~s; the
    branching fraction to the two-body decay mode is $90.2\%$. The dashed line represents the background and the
    solid line represents the background plus dark matter signal.  Data from the following experiments are shown:
    PAMELA \cite{Adriani:2010rc} (solid dots), WiZard/CAPRICE \cite{Boezio:1997ec} ($\diamond$), and
    BESS \cite{Orito:1999re} ($\bigtriangleup$). \textit{Right panel}: The corresponding graph for the
    antiproton to proton ratio. Data from the following experiments are shown: PAMELA \cite{Adriani:2010rc} (solid dots), IMAX \cite{Mitchell:1996bi} ($\star$), CAPRICE \cite{Boezio:1997ec} ($\diamond$) and BESS \cite{Orito:1999re}
    ($\bigtriangleup$).}
    \label{fig:protonmu}
\end{figure}

\begin{figure}
    \centering
        \includegraphics[width=8cm]{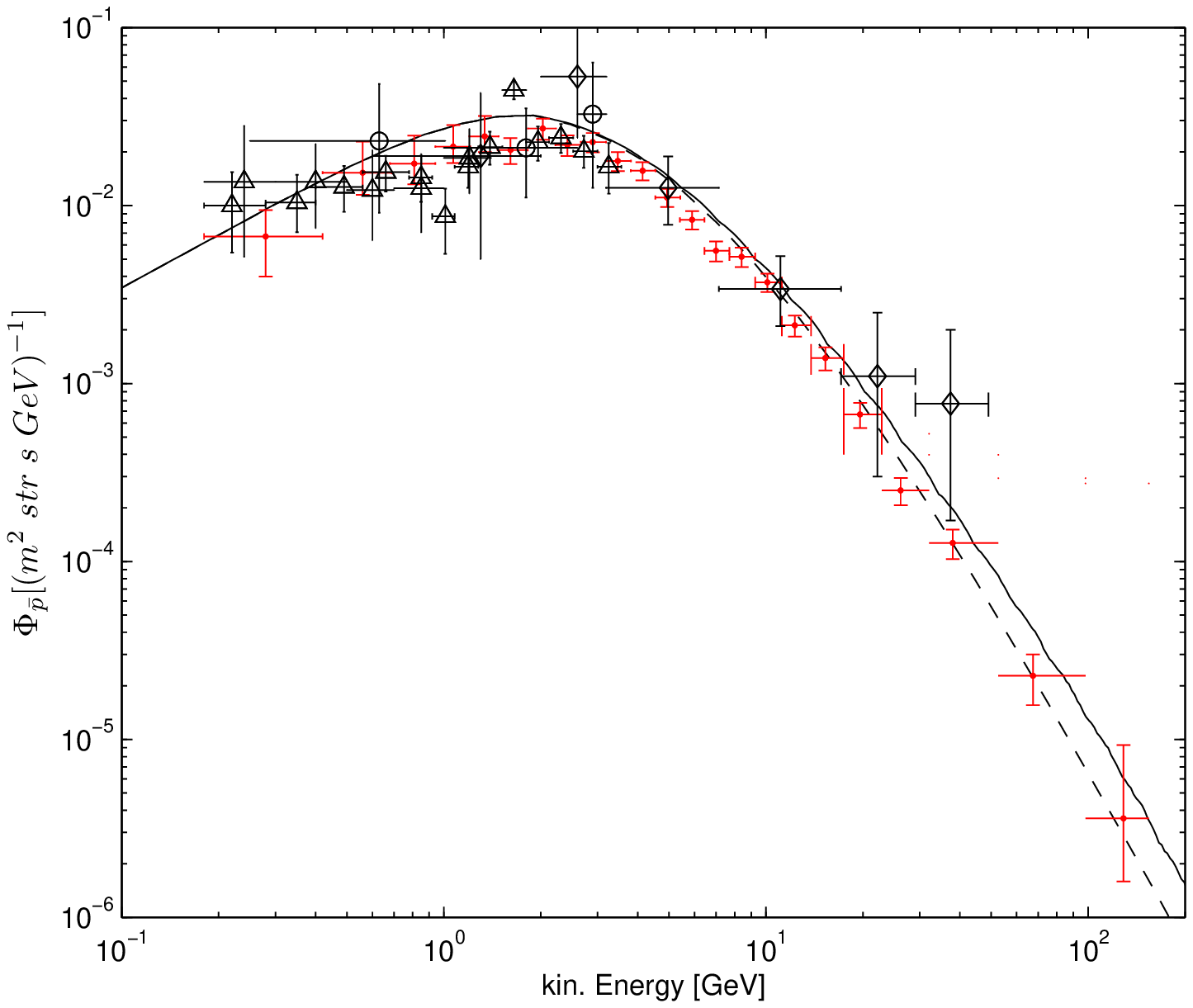}
        \includegraphics[width=8cm]{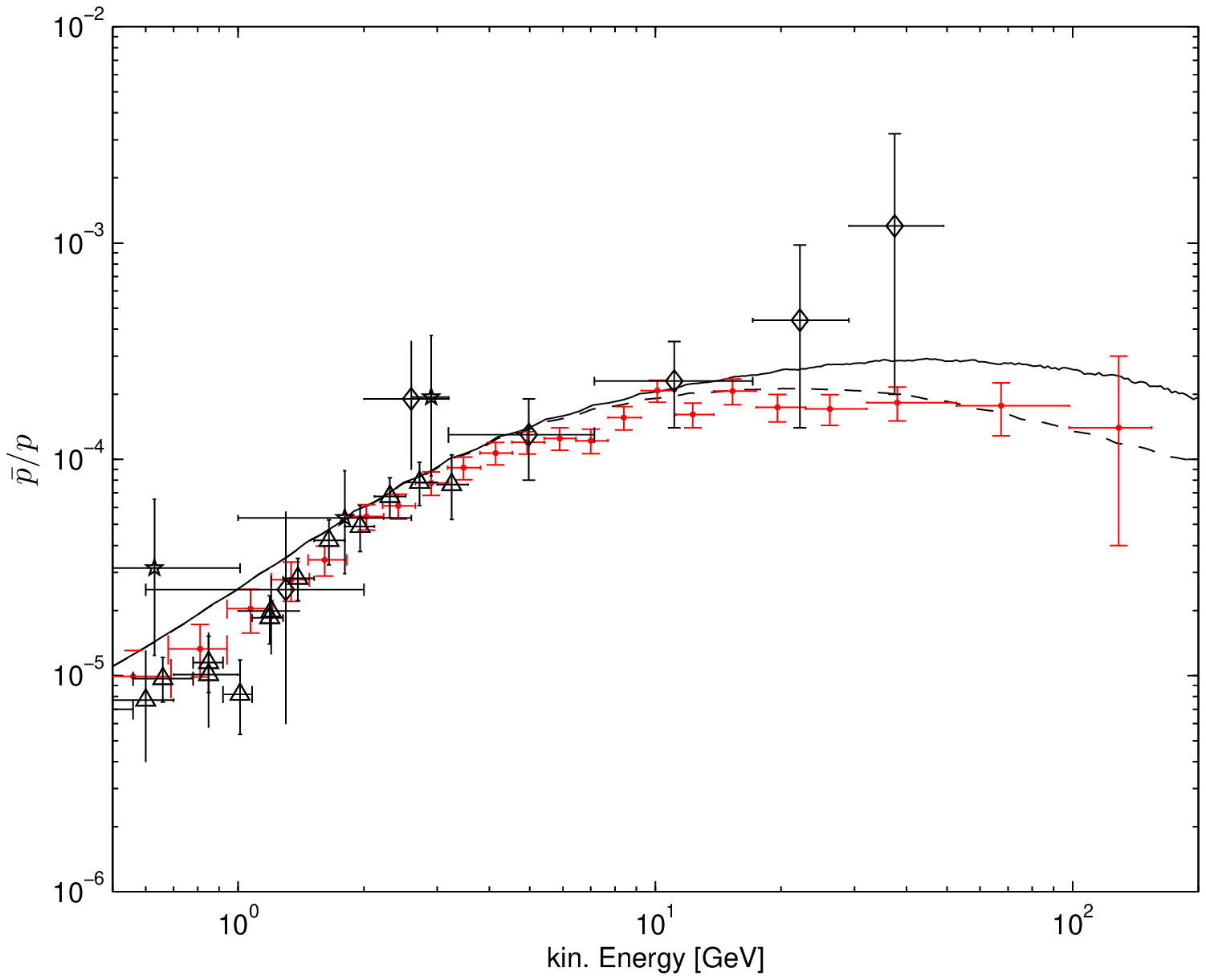}
    \caption{\textit{Left panel}: The antiproton flux for dark matter decaying into $\tau^-\tau^+$ and
    $\tau^-\tau^-h$. The dark matter mass is $5.0$~TeV and lifetime $9.0 \times 10^{25}$~s; the branching fraction
    to the two-body decay mode is 69.6\%. \textit{Right panel}: The corresponding graph for the antiproton to proton
    ratio.}
    \label{fig:protontau}
\end{figure}

Again assuming the MED propagation model~Ref. \cite{Delahaye:2007fr}, we compute the antiproton flux and
the antiproton to proton ratio for dark matter decays to $\mu^-\mu^+$ and $\mu^-\mu^+ h$, shown in
Fig.~\ref{fig:protonmu}, and for decays to $\tau^-\tau^+$ and $\tau^-\tau^- h$, shown in Fig.~\ref{fig:protontau}.
We see that in both cases, the antiproton excess above the predicted background curves is small and consistent
with the data shown from a variety of experiments.

\section{Relic Density and Direct Detection}

In this section, we show that the model we have presented can provide the correct dark matter relic
density while remaining consistent with the direct detections bounds.  The part of the Lagrangian that is relevant for computing the relic density, as well as the dark matter-nucleon elastic scattering cross section,
is the coupling between $\chi$ and standard model Higgs
\begin{equation}
\mathcal L \supset \lambda \chi^\dagger\chi H^\dagger H.
\end{equation}
In unitary gauge, this can be expanded
\begin{equation}
\mathcal L \supset \frac{\lambda}{2} \left( \chi^\dagger \chi\, h^2 + 2\, v_{ew}\,\chi^\dagger\chi\, h \right).
\label{eq:unints}
\end{equation}

\begin{figure}
    \centering
        \includegraphics[width=4cm]{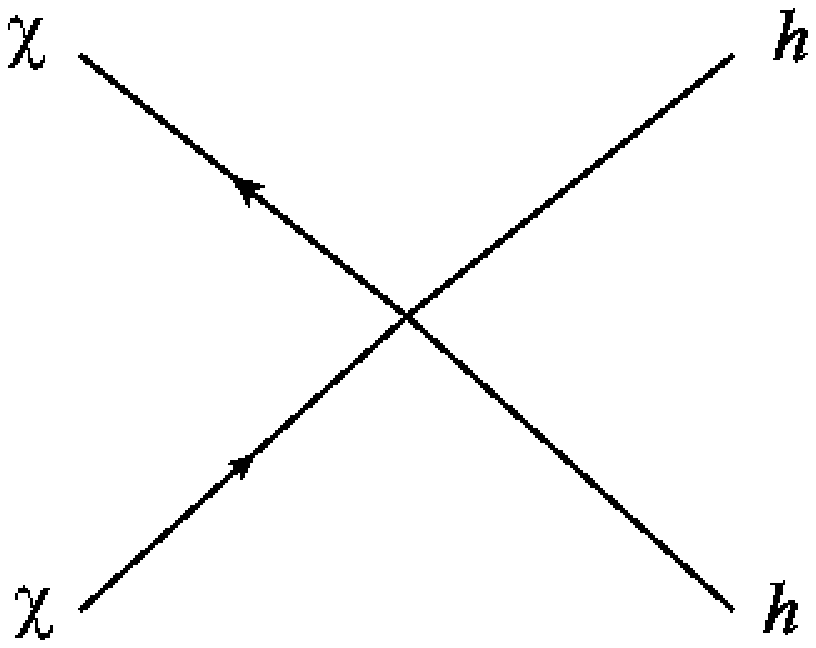}
        \includegraphics[width=5.7cm]{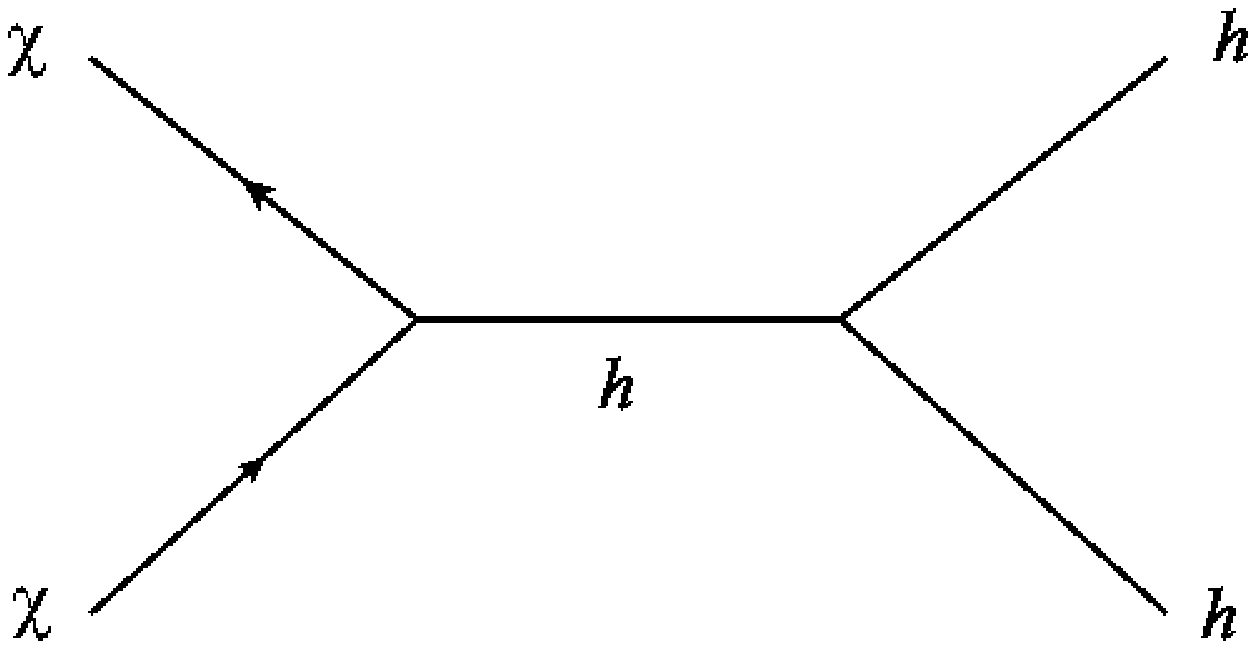}

        \includegraphics[width=4cm]{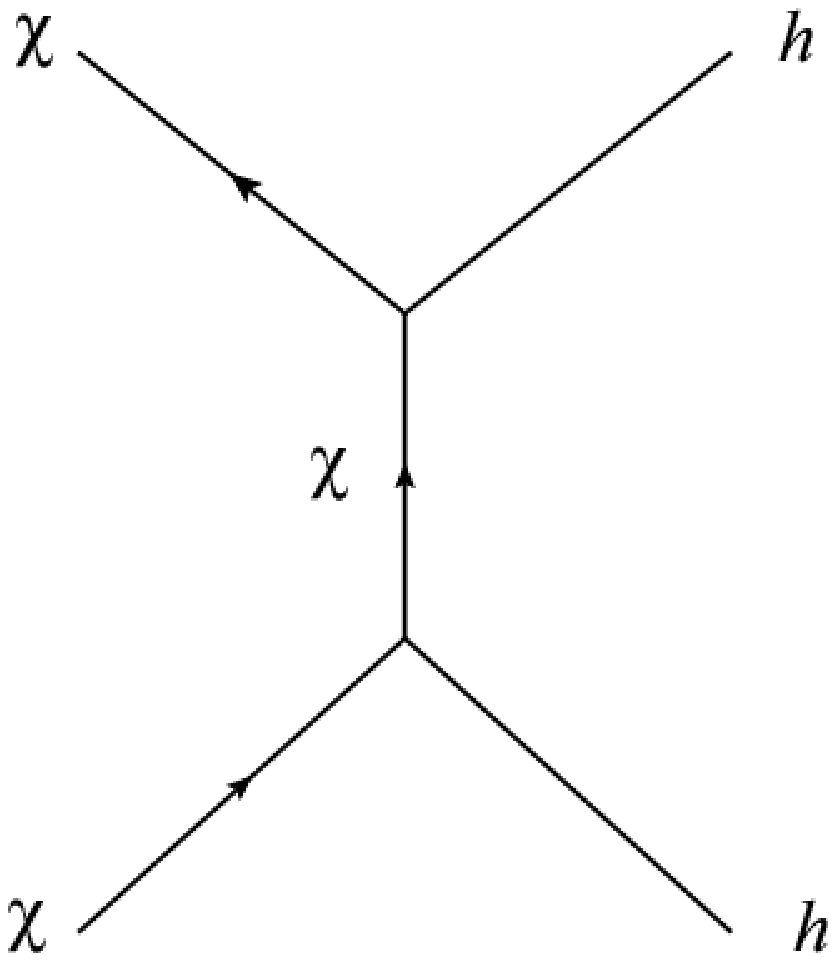}
        \includegraphics[width=4cm]{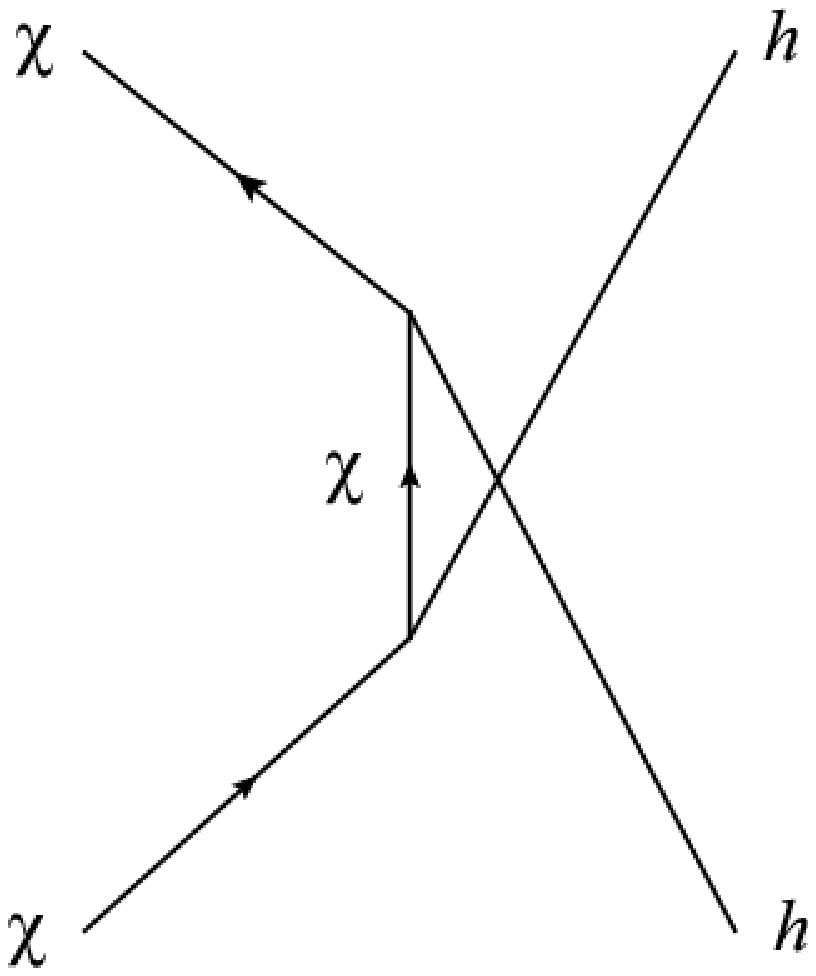}

        \includegraphics[width=5.5cm]{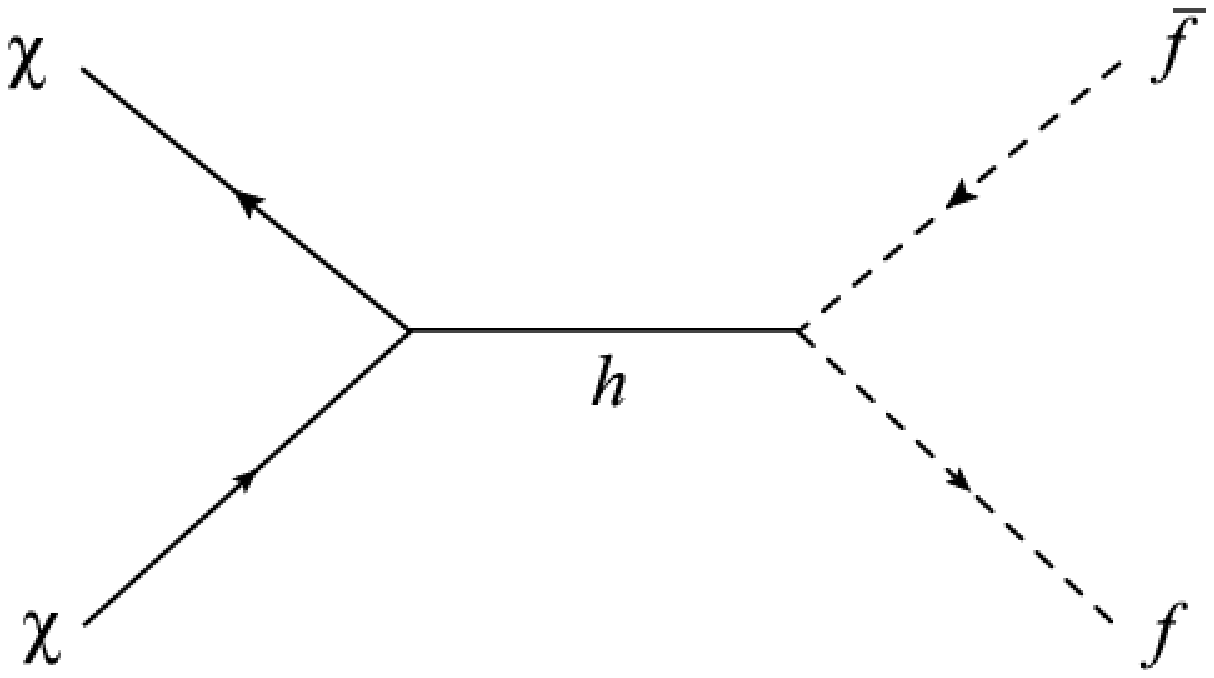}
        \includegraphics[width=6.2cm]{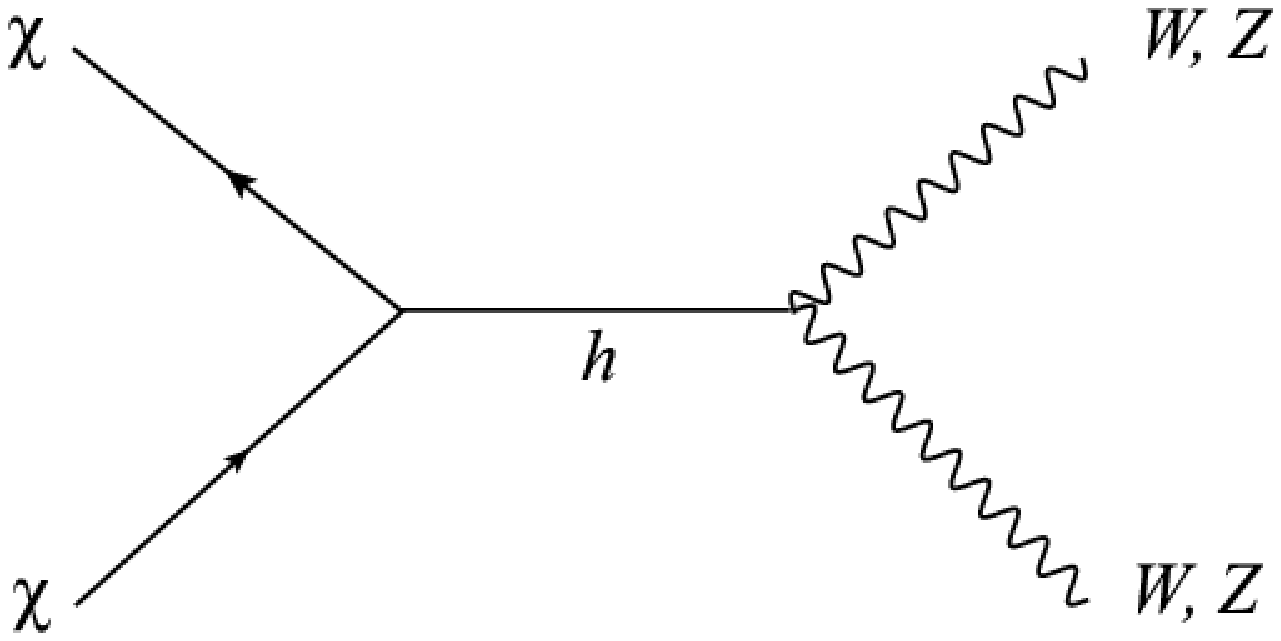}
    \caption{Dark matter annihilation diagrams.}
    \label{fig:annihilation}
\end{figure}
As a consequence of Eq.~(\ref{eq:unints}), $\chi$ and $\overline{\chi}$ pairs may annihilate into a
variety of standard model particles.  The leading diagrams are shown in Fig.~\ref{fig:annihilation}. The
cross section for annihilations into fermions is given by
\begin{equation} \label{eq:interaction}
\sigma_{\chi \bar\chi \rightarrow f\bar f} = \frac{N_c}{8\pi}\frac{\lambda^2m_f^2}{s\,(s-m_h^2)^2}\sqrt{\frac{\left(s- 4m_f^2\right)^3}{s-4m_\chi^2}},
\end{equation}
where $N_c$ is the number of fermion colors ($N_c=1$ for leptons and $N_c=3$ for quarks) and $m_f$ is the fermion mass. The cross sections for annihilations into $W$ and $Z$ bosons are given by
\begin{equation}
\sigma_{\chi \bar\chi \rightarrow ZZ} = \frac{\lambda^2}{8\pi}\frac{m_Z^4}{s\, (s-m_h^2)^2}(3-\frac{s}{m_Z^2}+\frac{s^2}{4m_Z^4})\sqrt{\frac{s-4m_Z^2}{s-4m_\chi^2}},
\end{equation}
\begin{equation}
\sigma_{\chi \bar \chi \rightarrow W^+W^-} = \frac{\lambda^2}{4\pi}\frac{m_W^4}{s\,
(s-m_h^2)^2}(3-\frac{s}{m_W^2}+\frac{s^2}{4m_W^4})\sqrt{\frac{s-4m_W^2}{s-4m_\chi^2}},
\end{equation}
where $m_W$ ($m_Z$) is the mass of $W$ ($Z$) boson. In the case where the dark matter annihilates into
a pair of standard model Higgs bosons, we can safely ignore the $t$- and $u$-channel diagrams since
the typical momenta are much smaller than $m_\chi$ at temperatures near freeze out.  Hence, the cross
section is given by
\begin{equation}
\sigma_{\chi \bar\chi \rightarrow hh} = \frac{\lambda^2}{32\pi \, s}  \sqrt{
\frac{s-4 m_h^2}{s-4m_\chi^2}} \left( 1 + \frac{6 m_h^2}{s-m_h^2} + \frac{9m_h^4}{(s-m_h^2)^2}  \right).
\end{equation}

The evolution of dark matter number density, $n_\chi$, is governed by the Boltzmann equation
\begin{equation}
\frac{dn_{\chi}}{dt}+3H(t) n_{\chi} = -\langle \sigma v \rangle [n_{\chi}^2-(n_{\chi}^{EQ})^2],
\end{equation}
where $H(t)$ is the Hubble parameter as a function of time and $n_\chi^{EQ}$ is the equilibrium number density.
The thermally-averaged annihilation cross section, $\langle \sigma v \rangle$, can be calculated by evaluating the integral \cite{Gondolo:1990dk}
\begin{equation}
\langle \sigma v \rangle =\frac{1}{8m_\chi^4TK_2^2(m_{\chi}/T)}\int_{4m_{\chi}^2}^\infty
(\sigma_{tot} )\, (s-4m_{\chi}^2)\sqrt{s} \, K_1(\sqrt{s}/T) \, ds \,\,\, ,
\end{equation}
where $\sigma_{tot}$ is the total annihilation cross section and the $K_i$ are modified Bessel functions of order $i$. We find the freeze out temperature, $T_f$, using the freeze-out condition~\cite{KolbTurner}
\begin{equation}
\frac{\Gamma}{H(t_F)} \equiv \frac{n_{\chi}^{EQ} \langle \sigma v \rangle}{H(t_F)} \approx 1 \,\, ,
\end{equation}
where equilibrium number density as a function of temperature is given by
\begin{equation}
n_{\chi}^{EQ} = \left(\frac{m_\chi T}{2 \pi}\right)^{3/2} e^{-m_\chi/T} \, .
\end{equation}
The Hubble parameter may be re-expressed as a function of temperature $T$
\begin{equation}
H=1.66 \,g_*^{1/2} \,T^2/m_{Pl} \, .
\end{equation}
where $g_*$ is the number of relativistic degrees of freedom and $m_{Pl}=1.22\times 10^{19}$~GeV is the Planck mass. It is customary to normalize the temperature with the dark matter mass, $x = m_\chi/T$.  For the points in parameter space discussed below, we found that the freeze out happens when $x_f \approx 28$. The present dark matter density can be calculated using the relation
\begin{equation}
\frac{1}{Y_0} = \frac{1}{Y_f} + \sqrt{\frac{\pi}{45}}m_{Pl} \: m_{\chi}\int_{x_f}^{x_0}
\frac{g_*^{1/2}}{x^2} \langle \sigma v \rangle \, dx \; ,
\end{equation}
where $Y$ is the ratio of number to entropy density and the subscript $0$ denotes the present time. The ratio of the dark matter relic density to the critical density $\rho_c$ is given by $\Omega_D = 2\, Y_0s_0m_{\chi}/\rho_c$, where $s_0$ is the present entropy density, or equivalently
\begin{equation}
\Omega_D h^2 \approx 5.6 \times 10^8\mbox{ GeV}^{-1} \, Y_0 \, m_\chi  \,\,\, .
\end{equation}
Note that the factor of $2$ included in the expression for $\Omega_D$ takes into account the contribution
from $\chi$ particles and $\bar\chi$ antiparticles.

In the case $m_\chi=2.5$~TeV, we find numerically that the dark matter-Higgs coupling $\lambda = 0.9$
in order that $\Omega_D h^2 = 0.1$.  For $m_\chi=5$~TeV, we find $\lambda=1.8$.  These order-one couplings
are perturbative.  One should keep in mind that the physics responsible for dark matter annihilations is
not directly linked to the mechanism that we have proposed to account for dark matter decay; other contributions to the total annihilation cross section can easily be arranged.  For example, if the Higgs sector includes
mixing with a gauge singlet scalar $S$ such that there is a scalar mass eigenstate near $2 m_\chi$, then the annihilation through the $s$-channel exchange of this state can lead to a resonantly enhanced annihilation
channel, as in the model of Ref.~\cite{CEP}. In this case, the correct relic density could be obtained
for smaller $\lambda$ than the values quoted above.

Finally, we confirm that the model does not conflict with bounds from searches for dark matter-nuclear recoil.
In this case, the most relevant contribution comes from the interaction between the dark matter and quarks mediated by a $t$-channel Higgs exchange. The effective Lagrangian is given by
\begin{equation}
\mathcal L = - \frac{\lambda \: m_q}{m_h^2} \chi^\dagger \chi \bar q q.
\end{equation}
Following Refs~\cite{McDonald:1993ex,Ellis:2000ds}, we can write an effective interaction between the
nucleons and dark matter,
\begin{equation}
\mathcal L = -(f_p \chi^\dagger \chi \overline{p} \, p + f_n \chi^\dagger \chi \overline{n} \, n) \,,
\end{equation}
where $f_N  = m_N \mathcal A_N \lambda / m_h^2$, for $N=p$ or $n$.  The coefficient $\mathcal A_N$ can be
evaluated using the results of Ref.~\cite{Ellis:2000ds};  numerically, one finds
$f_p \approx f_n \approx \mathcal A_N m_N \lambda / m_h^2$ with $\mathcal A_N \approx 0.35$.  Given the effective dark matter-nucleon interaction, we find that the spin-independent cross section is given by
\begin{equation}
\sigma_{SI} = \frac{\lambda^2 \mathcal A_N^2}{4 \pi}\frac{m_N^4}{m_h^4(m_\chi+m_N)^2}.
\end{equation}
For both of the cases discussed earlier, $(m_\chi=2.5\mbox{ TeV}, \, \lambda=0.9)$ and $(m_\chi=5\mbox{ TeV}, \, \lambda=1.8)$, we find $\sigma_{SI} \sim \mathcal O(10^{-45}) \textrm{ cm}^2$.  This is two orders of magnitude smaller than the strongest bounds, from CDMS~\cite{Ahmed:2009zw}, which range from $\sim 2 \times 10^{-43}$~cm$^2$ at $m_\chi=1$~TeV to $2 \times 10^{-42}$~cm$^2$ at $m_\chi=10$~TeV.

\section{Conclusions}
Models of decaying dark matter require a plausible origin for the
higher-dimension operators that lead to dark matter decays. The data from
cosmic ray experiments like PAMELA and Fermi-LAT require that
these operators involve lepton fields preferentially.  We have shown how
the desired higher-dimension operators may originate from Planck-suppressed couplings
between a TeV-scale scalar dark matter particle $\chi$ and vector-like states
at a mass scale $M$ that is intermediate between the weak and Planck scales.  The
vector-like sector has the structure of a Froggatt-Nielsen model:  charged lepton Yukawa
couplings arise only after these states are integrated out and a discrete gauged Abelian
flavor symmetry is broken.  Couplings between $\chi$ and the standard model gauge-invariant
combination $\bar L_L H e_R$ are then also generated, with coefficients of order
$\langle \phi \rangle^2/(M_*^2\,M)$, where $\langle \phi \rangle$ is the scale at which the flavor
symmetry is broken.  Taking $M$ and $\langle \phi \rangle$  near the geometric mean
of the reduced Planck scale and the weak scale, $O(10^{10})$~GeV, leads to the desired dark matter
lifetime. Neutrino masses can be generated via a conventional see-saw mechanism with the mass scale of
right-handed neutrinos also near $M$.  We pointed out that the symmetry structure of our model leads to
an overall suppression factor multiplying the charged lepton Yukawa matrix, but does not constrain
the standard model Yukawa textures otherwise.  Hence, our framework is potentially compatible with
a wide range of possible solutions to the more general problem of quark and lepton flavor in the
standard model.

We presented the necessary PYTHIA simulations to confirm that our model can account
for the anomalies observed in the cosmic ray experiments discussed earlier.  The leading
contribution to the primary cosmic ray  electron and positron flux in our model comes from
two-body decays, in which the Higgs field is set equal to its vev in the operator described
above; the subleading three body decays, $\chi \rightarrow \ell^+ \ell^- h^0$, are also
possible. We have checked that these decay channels do not lead to an observable excess
in the spectrum of cosmic ray antiprotons, since the cosmic ray antiproton flux
is in agreement with astrophysical predictions.

Our model demonstrates that the desired lifetime and decay channels of TeV-scale scalar dark
matter candidate can be the consequence of renormalizable physics at an intermediate lepton
flavor scale and gravitational physics at $M_*$. This presents an alternative scenario to
the one in which dark matter decay is a consequence of physics at a unification scale
located somewhere between $M$ and $M_*$.

\begin{acknowledgments}
We thank Josh Erlich and Marc Sher for useful comments. This work was supported by the NSF under
Grant PHY-0757481.  In addition, C.D.C. gratefully acknowledges support from a
William \& Mary Plumeri Fellowship.
\end{acknowledgments}


\end{document}